# Casimir self-energy of a free electron

**Allan Rosencwaig***

*Arist Instruments, Inc.*
*Fremont, CA 94538*

## Abstract

We derive the electromagnetic self-energy and the radiative correction to the gyromagnetic ratio of a free electron using a Casimir energy approach. This method provides an attractive and straightforward physical basis for the renormalization process.
.

* E-mail address: allan@aristinst.com

It has long been thought that the electromagnetic self-energy of the electron gives rise to the observed rest mass energy. The classical electrostatic self-energy, $e^2/r$, diverges as $1/r$ where $r$ is the electron radius and a classical radius, $a_0 = e^2/mc^2$, is defined as the radius at which the electrostatic energy is equal to the rest mass energy. Quantum electrodynamic (QED) calculations produce an electromagnetic self-energy that is also divergent in $1/r$ but the divergence is now logarithmic. The QED divergence is removed by renormalization processes which mathematically subtract two infinite terms in order to arrive at the measured rest mass energy.

We recently showed that we can obtain good agreement with QED calculations of the electromagnetic self-energy, or the one-photon radiative correction, for an electron in a hydrogen orbital using a Casimir energy approach [1]. In this approach, the self-energy is obtained by taking into account the probability of a single electron interacting with a virtual photon though the electromagnetic scattering cross-section and the probability that the electron is present at a radius $r$ in the atomic orbital through the appropriate hydrogen wavefunction. By using the Klein-Nishina formula for the scattering cross-section we make this treatment relativistic and therefore potentially applicable to the problem of the self-energy of a free electron. One advantage of such an approach is that the Casimir energy is the difference between infinitely large vacuum or zero-point energies of the quantum electromagnetic field inside and outside some configuration boundaries. Thus, the Casimir energy approach inherently incorporates a straightforward and attractive physical basis for the renormalization process.

As in our previous paper, we will make use of Klich's contour integral result for the Casimir energy of a macroscopic fully-conducting hollow sphere or shell of radius $r$ [2].

$$U_C = \frac{\hbar c}{4\pi} \int_0^\infty G(r,z)\,dz \qquad (1)$$

Where $\hbar$ is the reduced Planck constant, $c$ is the speed of light, $z = 1/\lambda$ and,

$$G(r,z) = e^{-4rz}\left(1 + 4rz + 4r^2z^2\right) \qquad (2)$$

Since in our case the shell is composed of a single electron, we take into account the probability that a virtual photon will interact with this single electron by adding into Eqn. (1) the total electromagnetic scattering cross-section, $\sigma_T$.

$$U_C = \frac{\hbar c}{4\pi r^2} \int_0^\infty \sigma_T(z) G(r,z)\,dz \qquad (3)$$

In Eqn. (3), the $r^2$ in the denominator normalizes the total scattering cross-section to the size of the shell. Using the well-known Klein-Nishina formula for the differential scattering cross-section,

$$\frac{d\sigma}{d\Omega} = \frac{1}{2}a_0^2 \left\{ \frac{1}{[1+\varepsilon(1-\cos\vartheta)]} - \frac{\sin^2\vartheta}{[1+\varepsilon(1-\cos\vartheta)]^2} + \frac{1}{[1+\varepsilon(1-\cos\vartheta)]^3} \right\} \quad (4)$$

$$\varepsilon = \frac{\hbar\omega}{mc^2}$$

where $\hbar\omega$ is the photon energy, $m$ is the electron mass, and $\vartheta$ is the scattering angle, we find that $\sigma_T$ can be written as,

$$\sigma_T = 2\pi a_0^2 f(\varepsilon) \quad (5)$$

with

$$f(\varepsilon) = \frac{(2+6\varepsilon+\varepsilon^2)(1+\varepsilon)^2 - (2+6\varepsilon+5\varepsilon^2) + 2(1+\varepsilon)^2(-2-2\varepsilon+\varepsilon^2)\ln(1+\varepsilon)}{2\varepsilon^3(1+\varepsilon)^2} \quad (6)$$

At any radius $r$, the largest $\lambda$ that can be supported by the shell is $\lambda = 2r$. This is true even though at shorter wavelengths the modes within the shell are circumferential whispering gallery modes where $\lambda = n(2\pi r)$ and $n$ is an integer. For the case of a free electron $r \ll \lambda_c$ where $\lambda_c$ is the electron Compton wavelength $= 2\pi\hbar/mc$. Thus $\lambda \ll \lambda_c$ and we then have $\varepsilon \gg 1$ for all $r$. For this high-energy region,

$$f(\varepsilon) \approx \frac{1}{\varepsilon}\left\{\frac{1}{2} + \ln(\varepsilon)\right\} \quad (7)$$

Setting $y = r/\lambda_c$ and since $\varepsilon = \lambda_c/\lambda$, the function $G(r,z)$ can be written as,

$$G(r,z) = G(y,\varepsilon) = e^{-4y\varepsilon}\left(1 + 4y\varepsilon + 4y^2\varepsilon^2\right) \quad (8)$$

The Casimir energy for a shell of radius $r$ that has only one electron in that shell would then be,

$$U_C = \frac{1}{2}a_0^2 \frac{\hbar c}{\lambda_c r^2} \int_{\varepsilon_{min}}^{\infty} f(\varepsilon) G(y,\varepsilon) d\varepsilon \quad (9)$$

There is a lower bound $> 0$ to the integral in Eqn. (9) since there is an upper bound on $r$ which is the radius of the electron. For any $r$,

$$\varepsilon_{min} = \frac{\lambda_c}{\lambda_{max}} = \frac{\lambda_c}{2r} \quad (10)$$

Unfortunately, the integral in Eqn. (9) is not readily performed. However, noting that in Eqn. (7), $f(\varepsilon)$ varies primarily as $1/\varepsilon$ and that $\ln(\varepsilon)$ changes relatively slowly, we can make the following approximations:
1). we can set the upper bound on the integral to $1000\varepsilon_{min}$, since the contribution of energies higher than that will be small;
2). we can set $\ln(\varepsilon)$ to have a fixed value at the logarithmic mid-point of the integration range, or $\ln(\varepsilon) = \ln(31.6\varepsilon_{min})$.

Thus, we can define an approximate $f(\varepsilon)$ as,

$$f_1(\varepsilon) = \frac{1}{\varepsilon}\left\{\frac{1}{2} + \ln(31.6\varepsilon_{min})\right\} \qquad (11)$$

The Casimir energy for a one-electron shell of radius $r$ is then,

$$U_C = \frac{\hbar c}{2}\frac{a_0^2}{r^2}\frac{1}{\lambda_c}\int_{\varepsilon_{min}}^{1000\varepsilon_{min}} f_1(\varepsilon)G(y,\varepsilon)d\varepsilon = \frac{\hbar c}{2}\frac{a_0^2}{r^2}\frac{1}{\lambda_c}h(y)$$

$$h(y) = 0.9315 + 0.2857\ln\left(\frac{1}{y}\right) \qquad (12)$$

First let us consider the electron as a shell of fixed radius $a_o$. Then,

$$U_C(shell) = \frac{\hbar c}{2\lambda_c}\left\{0.9315 + 0.2857\ln\left(\frac{2\pi}{\alpha}\right)\right\} = 0.23mc^2 \qquad (13)$$

since $\dfrac{\hbar c}{\lambda_c} = \dfrac{mc^2}{2\pi}$.

Now let us consider the electron as a sphere of uniform charge density and with radius $a_o$. Then,

$$U_C(sphere) = \frac{\hbar c}{2}\frac{a_0^2}{\lambda_c}\frac{3}{a_0^3}\int_0^{a_0} h(y)dr = \frac{3}{2}\frac{\hbar c}{a_0}\int_{y_{min}\to 0}^{a_0/\lambda_c} h(y)dy = \frac{3}{2}\frac{mc^2}{\alpha}(0.003657) = 0.75mc^2 \qquad (14)$$

Although both the shell model and the sphere model of the electron result in a Casimir self-energy close to the rest mass energy of the electron, the sphere model gives a better result. Considering the approximations made in deriving $f_1(\varepsilon)$, these results strongly indicate that the rest mass energy of the electron is indeed the electromagnetic self-energy arising from the interactions of the electron charge with the virtual photons of the vacuum quantum electromagnetic field.

It would also be instructive to calculate the angular momentum $J_C$ associated with the Casimir self-energy for the two models of the electron. Since this angular momentum arises from the interaction of the electron with the virtual photons, it will correspond to the radiative correction, (g-2)/2 to the anomalous gyromagnetic ratio of the electron's magnetic moment. Using Eqn. (12), the angular momentum for a shell of radius $r$ will be given by,

$$J_C = \frac{U_C}{\omega} = \frac{\hbar c}{2} \frac{a_0^2}{r^2} \frac{1}{\lambda_c \omega_c} \int_{\varepsilon_{min}}^{1000\varepsilon_{min}} f_1(\varepsilon) G(y,\varepsilon) \frac{d\varepsilon}{\varepsilon} = \frac{\hbar}{2} \frac{a_0^2}{r^2} \frac{1}{2\pi} j(y)$$

$$j(y) = 0.406 y \left\{ 3.26 + \ln\left(\frac{1}{y}\right) \right\}$$

(15)

Here $\omega$ is the photon frequency and $\omega_c$ the Compton frequency of the electron.

For the shell model where $r = a_o$, the Casimir angular momentum is,

$$J_C(shell) = \frac{\hbar}{4\pi} j\left(\frac{\alpha}{2\pi}\right) = \frac{\hbar}{4\pi}(0.004725) \qquad (16)$$

For the sphere model with $r = a_o$, the angular momentum is,

$$J_C(sphere) = \frac{3}{4\pi} \frac{\hbar \lambda_c}{a_0} \int_{y_{min} \to 0}^{a_0/\lambda_c} j(y)dy = \frac{3}{2} \frac{\hbar}{\alpha}(2.882 \times 10^{-6}) \qquad (17)$$

We thus get for the radiative correction to the gyromagnetic ratio,

$$\frac{g-2}{2}(shell) = 0.376 \times 10^{-3}$$

$$\frac{g-2}{2}(sphere) = 0.592 \times 10^{-3}$$

(18)

This compares to the QED value [3] of $\frac{g-2}{2}(QED) = 1.159 \times 10^{-3}$ which is in excellent agreement with experiment [4]. The agreement between the Casimir values of the radiative correction to the gyromagnetic ratio with the QED value is quite good considering the approximations made. Taking into account both the Casimir self-energy and the Casimir angular momentum, it would appear that the spherical model of the electron provides the better results.

We conclude that a Casimir approach to evaluating the self-energy of the electron strongly indicates that the rest mass energy of the electron comes from the electromagnetic energy arising from the interactions of the electron charge with the virtual photons of the vacuum quantum electromagnetic field. Even with the approximations used in this treatment, a model of an electron as a sphere of uniform charge density with a radius equal to the classical radius $a_o$, provides a Casimir self-energy of $0.75mc^2$, and a radiative correction to the gyromagnetic ratio within a factor of 2 of the QED value. In addition, as noted before, this Casimir approach provides an attractive physical basis for the renormalization process.

We are left with the puzzle that high-energy scattering experiments indicate that the electron is a point particle with no size. We can reconcile these experimental results with our Casimir model by realizing that the Casimir electron sphere is not necessarily the physical shape of an electron but rather that it defines the shape of a cavity around the electron that confines to some extent the quantum electromagnetic field. In a manner similar to the creation of a local curvature or cavity in the gravitational configuration space (ie spacetime itself) around a point mass, a point charge may cause a similar deformation or cavity in the local electromagnetic configuration space.

## **Acknowledgements**

The author would like to thank L. Kofman and J.R. Bond for their helpful comments.